# The Chicxulub Impactor: Comet or Asteroid?

**Steve Desch, Alan Jackson, Jessica Noviello, and Ariel Anbar assess the evidence for what type of object impacted the Earth and triggered the end-Cretaceous extinction, and suggest best practices for writing and reviewing interdisciplinary papers.**

The discovery by Alvarez et al. (1980) of significant iridium (Ir) in the global K-Pg boundary clay layer at Gubbio, Italy has revolutionized our understanding of the how the geology of Earth is affected by the astronomical goings-on in its neighborhood. Extrapolating from their Ir data, they inferred a 6.6 km-diameter, carbonaceous chondrite-like asteroid as an impactor. Soon after that finding, an Oort cloud comet was suggested as an impactor (Rampino and Stothers, 1984; Davis et al. 1984), and the idea that tidal disruption of comets passing the Sun could increase the impact rate at Earth was suggested (Bailey et al. 1992). The question of whether the Chicxulub impactor was an asteroid or a comet has been asked for nearly four decades. Since the identification of the Chicxulub crater (Hildebrand et al. 1991), the diameter of the impactor is more commonly accepted to be 10 km (e.g., Brittan 1997), but subsequent research has not undermined the logic used by Alvarez et al. (1980). The broad consensus is in favor of an asteroid impactor.

Against this backdrop, Siraj and Loeb (2021) have recently calculated that ~20% of Earth-crossing long-period comets (LPCs) would have been tidally disrupted by passage near the Sun, increasing the probability of one fragment impacting Earth by a factor of ~15. They presented geochemical evidence to assert that 100% of comets, but only 10% of Earth-crossing main belt asteroids (MBAs) would match the composition of the Chicxulub impactor. Combining these, they argued that LPCs are ~5-14 times more likely than MBAs to be the Chicxulub impactor, which would overturn the consensus and decisively favor a comet impactor.

The purpose of this article is to assess and provide context for this conclusion by Siraj and Loeb (2021). We first critique their geochemical arguments, which ignore the Ir layer that is central to the debate, and which also conflate carbonaceous chondrites with specific types of carbonaceous chondrites, effectively applying a double standard favoring comets. In fact, careful consideration of the geochemical evidence strongly favors a CM or CR carbonaceous chondrite, and rules out a cometary impactor. We also show that their calculations of impact rates are very sensitive to the assumed number of fragments produced during tidal disruption, and that for likely values the impact rate of comet fragments is far less likely than collisions by asteroids. We close by discussing best practices for the publication of interdisciplinary manuscripts bridging the astronomical and geological sciences.

**The Compositional Evidence**

The main constraint discriminating between a comet and an asteroid has long been the same Ir content of the global clay layer at the K-Pg boundary from which Alvarez et al. (1980) first inferred an extraterrestrial impactor. This clay layer follows distinct and predictable thickness and compositional patterns that tend to correlate negatively with distance from the Chicxulub impact crater (Smit 1999; Claeys et al. 2002; Goderis 2013, 2021). The total mass of Ir is estimated to be $2.0 - 2.8 \times 10^{11}$ g (Artemieva and Morgan 2009). This is compared to the Ir delivered by asteroids or comets under different scenarios. The impactor diameter D consistent with the Chicxulub crater is either D=10 km for an asteroid, or D=7 km for a comet, which generally collides at higher velocity (Brittan 1997). Brittan (1997) demonstrated that a D=10 km carbonaceous chondrite-like asteroid would deliver ≈2.3 × $10^{11}$ g of Ir, very satisfactorily matching the requirement. In contrast, a D=7 km comet is estimated to deliver only ~0.1 × $10^{11}$ g of Ir, because it is smaller and half ice. On this basis alone, an asteroid is strongly favored as the impactor, and a comet is practically ruled out.

Siraj and Loeb ignored the Ir constraint, but do consider that the Chicxulub impactor had a composition like carbonaceous chondrites. In doing so, however, they missed the distinctions between different types of carbonaceous chondrites. Meteoriticists identify several such types: CV, CK, CO, CR, CM, and CI, plus the unusual CH and CB, each type having slightly different chemical and isotopic composition. When considering an asteroid impactor, Siraj and Loeb took the fraction of Earth-crossing MBAs that are C-type (spectrally associated with carbonaceous chondrites) to be ~30%, with ~40% of those being CM-like (Bottke et al. 2007), concluding that only ~10% of MBAs could provide a match to the impactor. Meanwhile, they considered 100% of comets to match carbonaceous chondrites, and did not make further distinctions. This double standard made comets appear ~10 times more likely than asteroids, but comets are at most a factor of ~2 times more likely to be carbonaceous chondrite. Moreover, a careful consideration reveals that the Chicxulub impactor was a particular type of carbonaceous chondrite not matched by comets at all.

Several lines of evidence point to the Chicxulub impactor being either a CM or CR carbonaceous chondrite. The first (noted by Siraj and Loeb) is a fossil meteorite recovered from marine sediments deposited at K-Pg time in the North Pacific Ocean, which is very likely to sample the impactor itself (Kyte 1998). Based on a metal and sulfide content ≈4-8vol%, an inferred matrix abundance ≈30-60vol%, and the presence of >200 μm inclusions, Kyte (1998) favored a carbonaceous chondrite of type CV, CO, or CR, and allowed for the possibility of CM type despite their lower abundances of opaque minerals. In contrast, CI chondrites have essentially no metal, are > 99vol% matrix, and can be excluded.

The second line of evidence (also cited by Siraj and Loeb) is the excess of the isotope $^{54}$Cr (up to $\varepsilon^{54}$Cr ≈ +1.0) that has been measured in the marine clay layer (Shukolyukov and Lugmair 1998; Trinquier et al. (2006). This excess is matched only by certain types of carbonaceous chondrites. CV, CO, or CK chondrites have $\varepsilon^{54}$Cr < 1.0, too low to cause this anomaly. Trinquier et al. (2006) calculate the implied mixing fractions of extraterrestrial material in the marine clay layer are roughly 6-19% if it is CM chondrite-like (CR, CH and CB would be similar), and ≈1.7-2.6% if it is CI chondrite-like. The mixing ratios of impactor to terrestrial materials in the marine clay layer are estimated by other means to be 6.5± 2.7% (Kyte et al. 1980) or 7.9 ± 3.8% (Ganapathy 1980), so CM (or CR) chondrites provide a very satisfactory match. A CI composition can be ruled out at the ~99% probability level.

A third line of evidence (not considered by Siraj and Loeb) comes from platinum-group elements (PGEs) in the marine clay layer. The ratios between Pd, Ir, Rh, Ru, and Pt, especially Rh/Ir ratios, strongly favor carbonaceous chondrites of type CM or CO (Goderis et al. 2013). Sufficient data are lacking, but CR chondrites appear consistent as well. But CI chondrites are ruled out by the PGE evidence (Goderis et al. 2013).

A fourth line of evidence (also not considered by Siraj and Loeb) comes from abundances of extraterrestrial amino acids in the K-Pg clay layer. Zhao and Bada (1989) measured the abundances of isovaline and α-amino-isobutyric acid (AIB), two amino acids rare on Earth but common in carbonaceous chondrites. The AIB/Ir mass ratios were inferred to be > 100 in the clay layer, roughly comparable to the levels found in CM2 and CR2 chondrites, which have ~700 ppb Ir (Wasson and Kallemeyn 1988) and ~5,000 - 50,000 ppb AIB (Glavin et al. 2010). The AIB:isovaline ratios in the clay layer, ≈ 2-4 (Zhao and Bada 1989), also are roughly consistent with the ratios, ≈ 2.0, in CM2 and CR2 chondrites (Glavin et al. (2010). Significantly, the AIB abundances of CI chondrites are < 1000 ppb (Glavin et al. 2010), and the *total* amino acid contents of CV, CK, CO, CB chondrites are < 1000 ppb (Elsila et al. 2016), so all other carbonaceous chondrites (except CH) can be excluded.

These lines of evidence are reviewed in **Table 1**. The PGE and amino acid data, combined with the $\varepsilon^{54}$Cr and fossil meteorite evidence, point strongly to the Chicxulub impactor having a carbonaceous chondrite composition of type CM or CR in particular. All other types, especially CI chondrites, can be ruled out. At first this constraint would seem very restrictive: CM and CR chondrites comprise only a few percent of intact meteorite falls. But intact falls represent a very small fraction, << 1%, of the total meteoritic material striking the Earth (Bland 1996). Micrometeorites collected in Antarctica are overwhelmingly associated with CM and CR chondrites (Engrand and Maurette 1998). All this suggests that CM and CR chondrites are perhaps more representative of asteroids reaching Earth but are simply underrepresented among intact meteorite falls. At any rate, the ~40% of carbonaceous chondrites that are of type CM should be considered a reasonable match to the Chicxulub impactor. The fraction of Earth-crossing asteroids that are C-type is closer to 50% (Morbidelli et al. 2020), so the fraction of MBAs striking Earth that would match the impactor's composition is not ~10%, but in fact at least 20%.

**Table 1**: Comparison of different carbonaceous chondrite types with geochemical and other constraints from the fossil meteorite, the $^{54}$Cr anomaly, platinum-group elements, and amino acid abundances in the K-Pg clay layer. Only CM and CR chondrites provide a match.

| Constraint | CV | CK | CO | CH | CB | CM | CR | CI |
|---|---|---|---|---|---|---|---|---|
| Fossil Meteorite | yes | ? | yes | no | no | yes? | yes | no |
| $\varepsilon^{54}$Cr | no | no | no | yes | yes | yes | yes | no |
| PGE | no | no | yes | no | no | yes | yes? | no |
| Amino acids | no | no | no | ? | no | yes? | yes | no |

While Siraj and Loeb demand an asteroid match a CM composition, they only demand that a comet match a carbonaceous chondrite composition generally, and point to the *Stardust* comet return sample having a 'carbonaceous chondrite' composition (Zolensky et al. 2008) to assert that 100% of comets do. But comets are only reasonably a match to CI chondrites. As reviewed by Campins and Swindle (1998), cometary materials must be identified with a meteorite type that is rare (< 10 in our collections), dark, weak and friable, with low density, containing anhydrous silicates, and no chondrules. Almost all of these criteria are uniquely met by CI chondrites; in contrast, all other known carbonaceous chondrites, including CM, violate most of these constraints. Gounelle et al. (2006) made a strong case that the CI chondrite Orgeuil, whose fall was observed in 1864, originated from a Jupiter family comet and that CI chondrites may represent cometary materials generally. The protoplanetary disk model of Desch et al. (2018) predicts that the known carbonaceous chondrites formed roughly at Jupiter's orbit, except for CI chondrites, which must originate beyond Saturn's orbit, where comets are expected to form. In truth, comets probably sample a mix of both known and unknown carbonaceous chondrites; but if one had to select a known type to represent comets, it would only be CI, and would not be CM.

If the Chicxulub impactor is identified only with carbonaceous chondrites, then ~50% of MBAs and perhaps 100% of LPCs would qualify as a match. For comparable impact rates, a comet would be perhaps ~2× as likely as an asteroid to be the impactor. But if the Chicxulub impactor must be identified with a CM or CR (but not CI) chondrite, and comets are identified with a CI (but not CM or CR) chondrite, then > 20% of MBAs but ≈0% of LPCs would qualify as a match. Siraj and Loeb only concluded that comets were ~10× more likely than asteroids because they conflated carbonaceous chondrites with specific meteorite types, and ignored the Ir evidence.

**Impact Rates**

Beyond the geochemical evidence, Siraj and Loeb calculated the impact rates of comets to suggest they would be plausible impactors where asteroids would not be. The case against asteroids is weak. They claimed the background impact rates of MBAs are too low to explain the Chicxulub impact event, and can be dismissed. Yet their first paragraph states both that Chicxulub was the largest impact in the last 250 Myr, and that impacts of MBAs its size (diameter D > 10 km) should occur with mean interval $t_{MBA}$ ≈ 350 Myr. Thus the likelihood of a Chicxulub-scale asteroid impactor over the last 250 Myr is > 50%. Just as comet disruption could increase the impact rate, so could collisional disruption of a larger asteroid. Bottke et al. (2007) hypothesized that the breakup of the asteroid 298 Baptistina might have enhanced the MBA impact flux by a factor of 2 over the last ~100 Myr. Siraj and Loeb cite relevant literature to dispute this, but this is a straw man argument. Impact by an asteroid is plausible, even likely, even if Baptistina is an unlikely source.

In contrast, collision with a cometary impactor is only probable if it is a fragment from a larger LPC. The impact rate of Chicxulub-scale comets must be extrapolated from the numbers of smaller comets, but assuming a cumulative size distribution power law with index $q$=2, a comet with diameter D=7 km strikes the Earth with mean interval 3800 Myr, making the probability < 7% that one has impacted in the last 250 Myr. To make the impact more probable, Siraj and Loeb calculated that 20% of LPCs hitting Earth would first pass through the Sun's Roche limit and be tidally disrupted. They cited the evidence of Shoemaker-Levy 9 (Walsh 2018) and crater chains on Ganymede and Callisto (Schenk et al. 1996) to argue that all LPCs should disrupt into a number $N$ of equal-sized fragments, increasing the probability of a fragment impacting, by a factor $E = 0.2 \times N \times (D / 7 \text{ km})^{1-q}$. However, only those progenitor comets with diameter D > $N^{1/3}$ × (7 km) would yield a Chicxulub-scale impact, so the enhancement factor is $E = 0.2 \times N \times (N)^{(1-q)/3}$. The mean interval between impacts by comet fragments capable of making a Chicxulub-sized crater is then $t_{frag}$ ≈ 3800 Myr / $E$.

Assuming the existence of sufficiently large progenitor comets, the impact rate of fragments is maximized by assuming the largest possible value of $N$. Siraj and Loeb implicitly assumed a value $N$= 630, yielding $E$ ≈ 15, so that $t_{frag}$ ≈ 260 Myr. This value of $N$ was not well justified, and appears to have been chosen so that a typical size of comet, D = 60 km, would break up into fragments with diameter 7 km, just sufficient to create the largest possible number of Chicxulub-scale impacts. Even so, the impact rate of comet fragments (1 every 260 Myr) would not significantly exceed that of asteroids (1 every 350 Myr). However, the examples of Shoemaker-Levy 9 and the crater chains strongly suggest that comets typically are tidally disrupted into a much smaller number of fragments, $N \approx 10\text{-}30$, in which case the mean interval between collisions with comet fragments is 1 every 2000 Myr. Despite the importance of the number of fragments, $N$, Siraj and Loeb did not set it as a free parameter and explore the sensitivity of their results to it or acknowledge this major uncertainty in their calculation.

**Assessment of Probabilities**

We calculate the relative probability of the Chicxulub impactor being a comet or an asteroid, parameterizing the uncertain value of $N$. The relative probability of a comet impacting is $P(\text{comet}) = [\,1 + t_{\text{frag}} / t_{\text{MBA}}\,]^{-1}$, and of an asteroid is $P(\text{asteroid}) = 1 - P(\text{comet})$. Without fragmentation, $P(\text{comet}) = 8\%$. For a plausible value, $N \approx 20$, the relative probability of a comet is 12% if $q=2$, the slope of the size-frequency distribution for dynamically hot Kuiper belt objects (KBOs); or 22% if $q=2.9$, appropriate for dynamically cold KBOs (Fraser et al. 2014). Only if $q=2$ and $N>600$ would the probability of a comet become comparable to the probability of an asteroid.

We then use Bayes's theorem to combine the above impact rate probabilities with the constraint that the impactor must at least have an unspecified carbonaceous chondrite composition:

$P(\text{comet}|\text{CC}) =$
$\quad P(\text{CC}|\text{comet}) \times P(\text{comet}) \times [\,P(\text{CC}|\text{comet}) \times P(\text{comet}) + P(\text{CC}|\text{asteroid}) \times P(\text{asteroid})\,]^{-1}$,

where $P(\text{CC}|\text{asteroid}) \approx 50\%$ is the probability of a carbonaceous chondrite composition given an asteroid impactor, and $P(\text{CC}|\text{comet}) \approx 100\%$. The probability of a comet impactor based on just the impact rates and then based on the impact rates and the need to be a carbonaceous chondrite are plotted in **Figure 1** as a function of $N$, the number of cometary fragments per breakup. The requirement that the impactor match a carbonaceous chondrite somewhat increases the likelihood the impactor was a comet (to 21% if $q=2$, $N=20$), but in general a comet still would be less likely unless $N>600$.

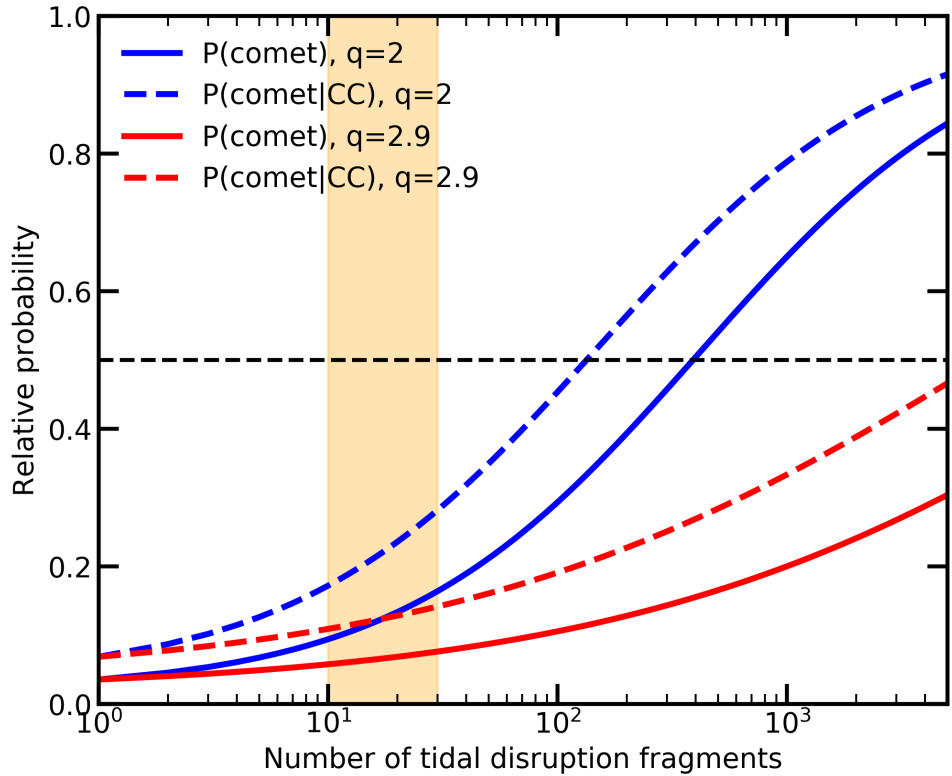

**Figure 1.** Relative probability of impacts (given an impact) by comets as opposed to an asteroid (solid curves), and the relative probability after imposing the constraint that the impactor was a carbonaceous chondrite of unspecified type (dashed lines), as a function of the number of fragments produced during a tidal disruption, $N$, for two different values of $q$, the exponent in the cometary size distribution. The value of $N$ is uncertain, but the example of Shoemaker-Levy 9 suggests $N \approx 10\text{-}30$ (shaded bar), in which case a comet is not favored (<50% probability, dashed horizontal line). Including the constraints that the impactor must match a CM or CR carbonaceous chondrite types and supply the Ir in the global clay layer, the probability of a comet is ≈0%.

Of course, if the geochemical criterion is that the impactor must match a CM or CR chondrite composition and/or must match the Ir anomaly, then the probability of an asteroid matching that exact composition is somewhat reduced but is still plausible, whereas a comet can be ruled out. The relative probability of a comet is 0%.

In summary, despite the claims made by Siraj and Loeb (2021), the case for an asteroid impactor is very strong. The enhancement in impact rates of comets due to tidal disruption is sensitive to the number of fragments generated per disruption, $N$, but they did not acknowledge its uncertainty or parameterize this input. Comet fragments are more likely impactors than asteroids only if the number of generated fragments is $N > 600$, the value arbitrarily and with little justification chosen by Siraj and Loeb, but far exceeding the values $N \approx 10\text{-}30$ suggested by observations of Shoemaker-Levy 9 and crater chains.

Siraj and Loeb (2021) claimed the likelihood of a comet is increased after imposing the constraint that the impactor had carbonaceous chondrite composition; but they applied a double standard by requiring the asteroid impactor to be a CM chondrite, but demanding comets be any particular type of carbonaceous chondrite. In fact, the impactor must be a CM or CR chondrite, and this makes asteroids plausible but rules out comets, which are only strongly associated with CI chondrites. Of course, the observed amount of Ir in the K-Pg boundary clay layer argues in favor of an asteroid but rules out a comet, but this key evidence was ignored entirely by Siraj and Loeb (2021).

**The Challenges of Interdisciplinary Science**

The nature of the Chicxulub impactor is an outstanding problem at the intersection of Earth science and astronomical sciences. Problems like these are of great general interest. As they allow researchers in one field to leverage the results of another, these sorts of investigations should be encouraged. But the fact that the work by Siraj and Loeb (2021) has so many flaws easily rectified by a quick review of the literature highlights many of the challenges to engaging in interdisciplinary science.

One challenge is that despite the greater need to synthesize the literature, there are fewer venues for doing so. While the constraints about which carbonaceous chondrite types match the impactor are all available, these are dispersed among a variety of journals (*Earth and Planetary Science Letters, Geochimica et Cosmochimica Acta, Meteoritics and Planetary Science, Nature, etc.*). It is difficult for an astrophysicist to become fluent in researching this entirely new field of literature, which samples more journals than in astrophysics.

Another challenge is the need to go beyond just accessing information from another field, to understanding the context and significance of that information. In the Internet era, it is easier than ever to learn of findings from another field. It is not as easy, but still possible, to learn the jargon of that field and know how the data were acquired. More difficult still, though, is appreciating the context of the information: What are that field's underlying, unspoken assumptions, and what information is missing? Often there are differences in scientific culture between fields about how they deal with uncertainty, or what constitutes a burden of proof. It is possible and rewarding to engage in interdisciplinary research, but it starts with opening dialogs with researchers in other fields, based on mutual respect and a lot of listening.

Interdisciplinary research also poses challenges to the peer review process itself.
It is difficult but necessary to find the needed range of reviewers for papers like the one by Siraj and Loeb (2021), bridging celestial mechanics a*nd* cratering statistics *and* meteoritics *and* geochemistry. That this paper was not reviewed by multiple scientists with expertise across the relevant fields is evidenced in ways both large and small.

For example, a reviewer familiar with geochemistry certainly would have demanded the authors at least *address* Ir. Siraj & Loeb cited Alvarez et al. (1980) for introducing the concept of an impactor, yet seemingly failed to recognize the significance of Ir, or that it is a stringent geochemical constraint key to diagnosing between a comet and an asteroid (Brittan 1997; Artemieva and Morgan 2009), ostensibly the main quest of the paper. It is one of many reasons, beyond impact rates, why asteroids have long been favored over comets, so to not even mention the important constraint of Ir in a paper seeking to revive the debate of comet vs. asteroid is a serious omission that any geochemist would have caught.  A referee familiar with meteoritics probably would be needed to catch the significant conflation of CM chondrites with all carbonaceous chondrites. But a reviewer with a background in astrophysics and the culture of how models deal with uncertainty, would be most likely to demand that the number of fragments be considered a free parameter, and the sensitivity of the results to that parameter explored. Each of these mistakes severely undercuts the authors' arguments, but only a rare single reviewer would have caught all of them.

Other issues are less substantive but equally telling that the paper was not reviewed by referees from different disciplines. Perhaps it wouldn't take an astrophysicist to catch the logical inconsistency that something happening once per 350 Myr was dismissed as too rare to happen once in 250 Myr. But only someone familiar with asteroids would have noticed the name "Baptistina" was misspelled repeatedly. A geologist would have complained that the defunct term "K-T" was used instead of "K-Pg" (Cretaceous-Paleogene). "K-Pg" has been standard since 2009, and "K-T" is discouraged by the International Commission on Stratigraphy. Likewise, a paleontologist would have objected to the title of the paper, which purports to explore the cause of the 'dinosaur extinction'. While it is commonly accepted that the Chicxulub impact is associated with and likely precipitated the end-Cretaceous mass extinction event that killed ~75% of all plant and animal species on land and in the oceans, not just dinosaurs (and, more precisely, just the non-avian dinosaurs), this remains an area of ongoing scholarship (e.g., Schulte et al. 2010; Chiarenza 2020), and at no point did the paper explore the 'dinosaur extinction'. The authors chose a title flashier than the more accurate 'origin of the Chicxulub impactor'; but it is not scientifically rigorous, and a geologist or paleontologist reviewer would have objected.

The solution to the problem of how to review interdisciplinary papers is for journal editors to find reviewers spanning all the disciplines pertinent to the paper, though this is easier said than done. Finding and accommodating multiple reviewers takes longer and is at cross purposes with making manuscripts "swiftly visible." Being deliberative and making a paper's conclusions precise is at cross purposes with making manuscripts "highly discoverable." But these practices must be applied to interdisciplinary papers to ensure they meet the scientific standards of all the fields involved, so that disciplines can build off each other's results.